\magnification=1200

{\bf A precise determination of angular momentum in the black hole candidate
GRO J1655-40}
\bigskip

Marek Artur Abramowicz and W{\l}odek Klu\'zniak
\smallskip
Institut d'Astrophysique de Paris, 98bis Boulevard Arago, F 75014 Paris,
France
\bigskip
\centerline{ABSTRACT}
\bigskip
We note that the recently discovered 450 Hz frequency in the X-ray flux of
the black hole candidate GRO J1655-40 is in a 3:2 ratio to
the previously known 300 Hz frequency of quasi-periodic oscillations (QPO)
in the same source. If the origin of high frequency QPOs in black hole
systems is a resonance between orbital and epicyclic motion of
accreting matter, as suggested previously, the angular momentum of the black
hole can be accurately determined, given its mass. We find that the
dimensionless angular momentum is in the range $0.2<j<0.65$
if the mass is in the (corresponding) range of 5.5 to 7.9 solar masses.  
\bigskip\bigskip

We have previously suggested [1] that ``twin''
kHz QPOs in accreting neutron stars
arise as a result of non-linear 1:2 or 1:3 resonance between
the radial epicyclic motion and the orbital motion of matter
in a nearly keplerian accretion disk, and we have noted that
the same phenomenon should also arise in accreting black holes,
where only a single high frequency had been observed.
Strohmayer now reports [2] the discovery of a second QPO
in GRO J1655-40, a well known black hole candidate in a low-mass X-ray
binary, with the mass of the compact X-ray source determined from
optical studies to be in the range $5.9<M/M_\odot<7.9$ [3].
There are time intervals, when both QPOs are present at the same time [2].

The two QPOs now known in the source occur at frequencies 300 Hz
and 450 Hz, i.e., in  a 2:3 ratio, strongly supporting the notion
of a resonance in the system. Of all rational ratios
only 1:2 and 1:3 resonances are capable of giving a 2:3 ratio of
frequencies. Specifically, if the lower frequency in the resonance is
$\omega$, and the higher frequency $\Omega$, the only two possibilities
are that $ \Omega=$ 300Hz and $\Omega+\omega=$ 450 Hz for the 1:2
resonance, or that 
$ \Omega=$ 450Hz and $\Omega-\omega=$ 300 Hz for the 1:3 resonance.

In the spirit of ref. [1], we consider
resonances between orbital and epicyclic motion in the Kerr metric.
 For the given mass range of the black hole, and for the frequencies
of 300 Hz and 450 Hz, we can exclude that orbital frequency
is in resonance with the epicyclic motion in
``vertical'' (meridional) direction.
Only resonance with radial epicyclic motion can give the
observed frequencies. So in the case of GRO J1655-40,
$\omega$ is the radial epicyclic frequency and $\Omega$ is the orbital
frequency for circular equatorial orbits in the Kerr metric.
The formulae for the frequencies can be found
e.g., in the review by Kato [4].

We find the following angular momentum range:

$0.2<j<0.6$ for the 1:2 resonance, and

$0.3<j<0.65$ for the 1:3 resonance.

Here $j=JG/(cM^2)=a$ is the Kerr parameter of the black hole,
i.e., its dimensionless angular momentum.

In both cases the upper limit is for $M=7.9M_\odot$,
and the lower limit is for $M=5.5M_\odot$. Thus, most of the uncertainty
in the determination of angular momentum comes from the mass uncertainty
and not from the choice of the resonance.

We conclude that the black hole in GRO J1655-40 is neither
a Schwarzschild nor a maximal Kerr black hole (contrary to some
reports). 
This is certainly expected [1]. The black hole is accreting matter
from a low-mass companion in a long lived binary system,
so it must have acquired considerable angular momentum in its
evolution. Because the companion is a low-mass star, the fractional
mass and angular momentum accreted by the black hole cannot be large.
Note that if the black hole mass is close to its ``preferred'' value
of $6M_\odot$, the Kerr parameter has a moderate value of about 0.3.

The source is also one of the Galactic microquasars,
so the presence of mass ejection in ``jets'' is not a signature
of a maximally rotating black hole.
 
It is a pleasure to thank Jean-Pierre Lasota and Eric Gourgoulhon
for helpful discussions.
\bigskip
REFERENCES

[1] Klu\'zniak, W.. Abramowicz, M.A. astro-ph/0105057

[2] Strohmayer, T.E. astro-ph/0104487

[3] Shahbaz, T. et al. 1999, MNRAS 306, 89

[4] Kato, S., 2001, PASJ 53, 1
\bye